\documentclass[12pt,preprint]{aastex}
\usepackage{graphicx}

\def\be{\begin{equation}}
\def\ee{\end{equation}}
\def\ba{\begin{eqnarray}}
\def\ea{\end{eqnarray}}
\newcommand{\msun}{\ifmmode\mbox{M}_{\odot}\else$\mbox{M}_{\odot}$\fi}
\newcommand{\rsun}{\ifmmode\mbox{R}_{\odot}\else$\mbox{R}_{\odot}$\fi}
\newcommand{\degrees}{\ifmmode^{\circ}\else$^{\circ}$\fi}
\newcommand{\degree}{\ifmmode^{\circ}\else$^{\circ}$\fi}
\newcommand{\amin}{\ifmmode^{\prime}\else$^{\prime}$\fi}
\newcommand{\asec}{\ifmmode^{\prime\prime}\else$^{\prime\prime}$\fi}
\newcommand{\etal}{et al.}
\newcommand{\gps}{\ensuremath{g_{\rm P1}}}
\newcommand{\rps}{\ensuremath{r_{\rm P1}}}
\newcommand{\ips}{\ensuremath{i_{\rm P1}}}
\newcommand{\zps}{\ensuremath{z_{\rm P1}}}
\newcommand{\yps}{\ensuremath{y_{\rm P1}}}
\def\psr{{\rm PSR J1048+2339}}

\shorttitle{Redback Pulsar J1048$+$2339}
\shortauthors{Deneva et al.}

\begin{document}

\title{Multiwavelength Observations of the Redback Millisecond Pulsar J1048$+$2339} 
\author{
J.~S.~Deneva$^{1,*}$, 
P.~S.~Ray$^{2}$,
F.~Camilo$^{3}$,
J.~P.~Halpern$^{3}$,
K.~Wood$^{2}$,
H.~T.~Cromartie$^{4}$,
E.~Ferrara$^{5}$,
M.~Kerr$^{6}$,
S.~M.~Ransom$^{7}$,
M.~T.~Wolff$^{2}$,
K.~C.~Chambers$^{8}$,
E.~A.~Magnier$^{8}$
}

\affil{$^{1}$National Research Council, resident at the Naval Research Laboratory, Washington, DC 20375, USA}
\affil{$^{2}$Naval Research Laboratory, Washington, DC 20375, USA}
\affil{$^{3}$Columbia Astrophysics Laboratory, Columbia University, New York, NY 10027, USA}
\affil{$^{4}$University of Virginia, Charlottesville, VA 22904, USA}
\affil{$^{5}$Goddard Space Flight Center, Greenbelt, MD, 20771, USA}
\affil{$^{6}$CSIRO Astronomy and Space Science, Marsfield NSW 2122, Australia}
\affil{$^{7}$National Radio Astronomy Observatory, Charlottesville, VA 22903, USA}
\affil{$^{8}$Institute for Astronomy, University of Hawaii, Honolulu, HI 96822, USA}

\begin{abstract}
We report on radio timing and multiwavelength observations of the
4.66~ms redback pulsar J1048+2339, which was discovered in an
Arecibo search targeting the {\em Fermi}-LAT source
3FGL~J1048.6+2338. Two years of timing 
allowed us to derive precise astrometric and orbital parameters for
the pulsar. PSR J1048+2339 is in a 6-hour binary, and exhibits radio
eclipses over half the orbital period and rapid orbital period variations. 
The companion has a minimum mass of 0.3 \msun, and we have identified a
$V \sim 20$ variable optical counterpart in data from several surveys.
The phasing of its $\sim 1$~mag modulation at the orbital period suggests
highly efficient and asymmetric heating by the pulsar wind, which may be
due to an intrabinary shock that is distorted near the companion,
or to the companion's magnetic field channeling the pulsar wind to specific
locations on its surface. 
We also present gamma-ray spectral analysis of the source and
preliminary results from searches for gamma-ray pulsations using the
radio ephemeris. 
\end{abstract}

\section{Introduction}

Data from the Large Area Telescope (LAT) on the {\em Fermi} Gamma-ray
Space Telescope have shown definitively that millisecond pulsars
(MSPs) are gamma-ray emitters similar to young pulsars \citep{Abdo09},
and more than a quarter of all known MSPs outside of globular clusters
were discovered in radio searches targeting {\em Fermi} unidentified
sources only within the past six years \citep{Ray12}. {\em Fermi} has
been especially successful in uncovering MSPs in black widow and
redback binary systems, where the pulsar's intense particle wind
gradually ablates the companion. The main difference between the two
classes is the mass and type of the companion. While in black widow
systems the companion is degenerate, with a mass of 0.01--0.05~\msun, redback pulsars have non-degenerate companions whose mass
is an order of magnitude larger \citep{Roberts11}. Both types are
believed to be transitional between accreting low-mass X-ray binaries (LMXBs),
rotation-powered binary MSPs, and isolated MSPs. Nineteen of 25 known
black widows and nine of 13 known redbacks in the Galactic field were
discovered in radio searches of {\em Fermi} unidentified sources. 

Black widows and redbacks are relatively rare objects because they are
selected against in undirected (``all sky'') surveys due to their
intrinsic properties (radio eclipses, high line-of-sight acceleration
from orbital motion) and because this stage in MSP evolution is short
compared to an MSP's lifetime. Consequently, there is much still to be
learned about the details of the transition between accreting and
rotation-powered MSPs (\citealt{Roberts11}, \citealt{Roberts13}). This
is especially true of redback pulsars, as they are the rarer and more
recently discovered type.

In this paper we present a detailed multiwavelength study of
PSR~J1048+2339, a redback radio pulsar found in a {\em Fermi} source
by \cite{Cromartie15}. The pulsar's prolonged radio eclipses around
superior conjunction as well as ``mini-eclipses'' at other orbital
phases are reminiscent of the redback class archetype, PSR~J1023+0038,
whose transition from LMXB to radio MSP was observed almost in real
time (\citealt{arc09}, \citealt{arc13}). In addition, PSR~J1048+2339
exhibits significant orbital period variations similar to those
observed in the redbacks PSR~J1023+0038, PSR~J2339$-$0533
\citep{ple15}, and the original black widow pulsar B1957+20
\citep{arz94}.

In Section~\ref{sec_radio} we present two years of radio timing
observations and polarimetry. In Section~\ref{sec_optical} we focus on
the optical properties of the system, and in Section~\ref{sec_gamma}
we discuss the gamma-ray counterpart.

\section{Radio Observations}\label{sec_radio}

PSR~J1048+2339 was discovered in a search at 327 MHz with the Arecibo telescope
of high-latitude {\em Fermi} unidentified sources from a preliminary version 
of the 3FGL {\em Fermi} LAT source catalog  \citep{Cromartie15}. 
The source
selection criteria were (1) non-variability in gamma-rays, (2)
pulsar-like gamma-ray spectrum, and (3) an error box 15\arcmin\ across
or smaller, corresponding to the diameter of the Arecibo beam at
327~MHz. The pulsar was discovered in a 15-minute observation targeting 
the {\em Fermi} source 3FGL~J1048.6+2338 on 2013 August 4.
PSR J1048+2339 has a pulse period of 4.66 ms and a 
dispersion measure (DM) of 16.65~pc~cm$^{-3}$.
According to the NE2001 model of
electron density in the Galaxy \citep{NE2001}, this DM corresponds to
a distance of 0.7~kpc.

We performed timing observations of PSR J1048+2339 at 26 epochs
between 2013 August 4 and 2015 September 20 using the Arecibo radio
telescope at 327~MHz. The observations used the PUPPI
backend in ``fast4k'' mode,
with 4096 channels and a 81.92~$\mu$s sampling time. Since the minimum
bandwidth provided by PUPPI, 100~MHz, is larger than the receiver
bandwidth of 68~MHz, only 2680 channels were recorded to disk,
corresponding to the part of the PUPPI band covering the receiver
bandwidth. After deriving a timing solution in 2015 July, 
we observed PSR~J1048+2339 at 1430~MHz at two epochs in coherent folding,
full-Stokes mode using PUPPI with a bandwidth of 700~MHz, 448 channels, and
a 64~$\mu$s sampling time. 

\begin{figure}
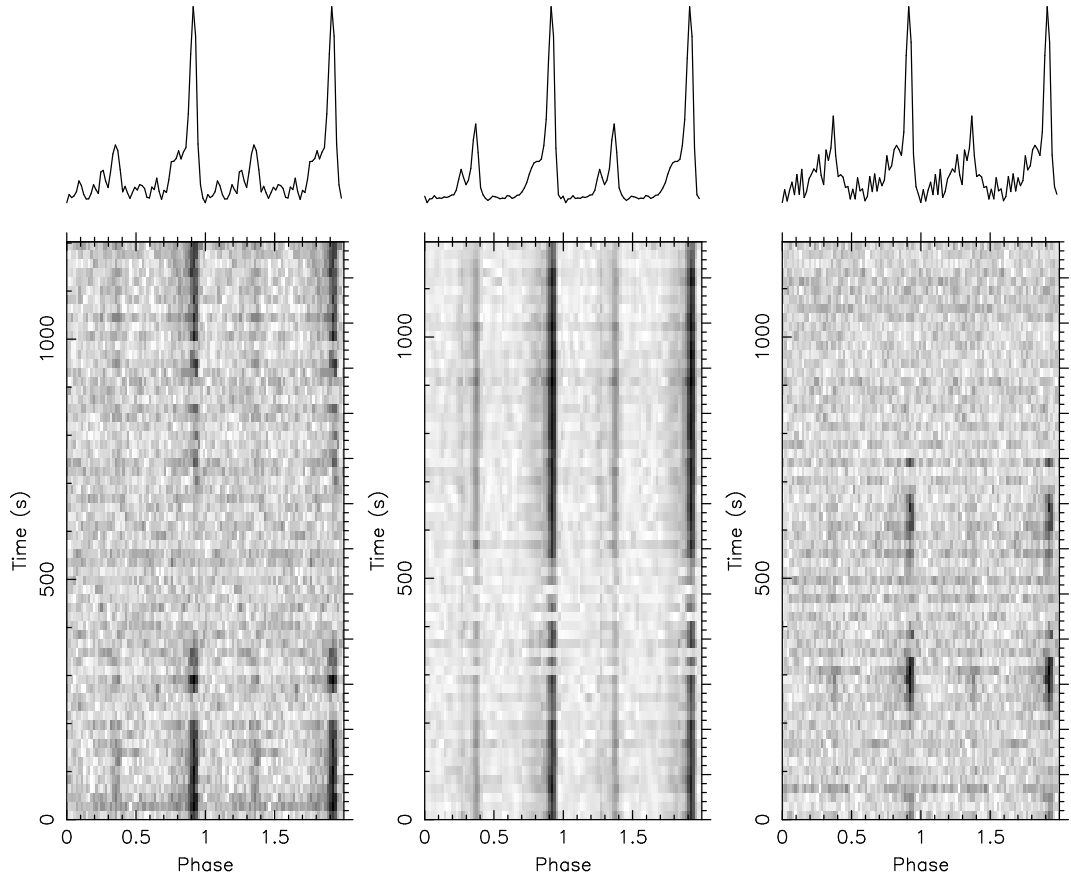

\begin{center}
\includegraphics[width=0.7\textwidth,angle=-90]{FIG1_left.ps}
\includegraphics[width=0.7\textwidth,angle=-90]{FIG1_middle.ps}
\includegraphics[width=0.7\textwidth,angle=-90]{FIG1_right.ps}
\caption{Folded pulse profiles and subintegration vs.\ pulse phase for three epochs.  Left: MJD~57084, $\phi_{\rm orb} = 0.70 - 0.76$. Middle: MJD~57075, $\phi_{\rm orb} = 0.84 - 0.89$. Right: MJD~57064, $\phi_{\rm orb} = 0.94 - 0.99$; eclipse ingress occurs at $\phi_{\rm orb} \sim 0.97$ ($t \sim 700$~s) during this observation. Two full rotations are shown in each panel. \label{fig_ingress}}
\end{center} 
\end{figure}

\begin{figure}
\begin{center}
\includegraphics[width=0.99\textwidth]{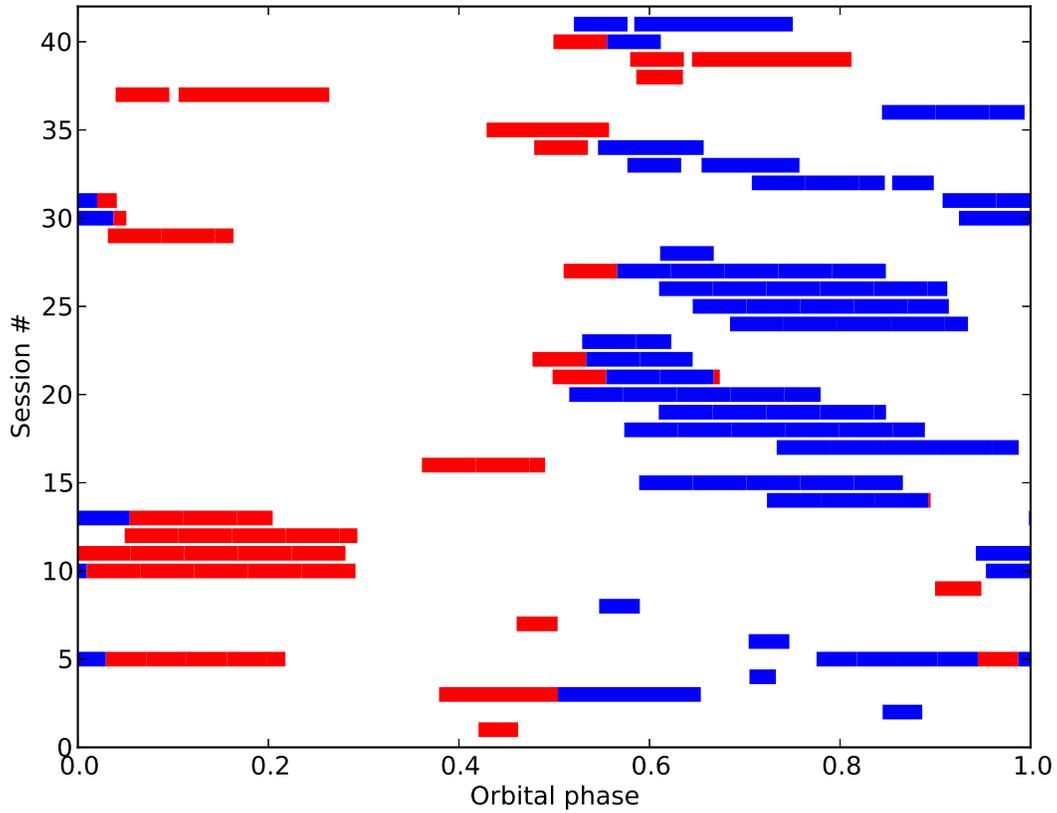}
\caption{Observations of PSR~J1048+2339 vs.\ orbital phase where the pulsar was detected for at least part of the integration time are shown in blue, and non-detections are shown in red. Sessions 3 and 4 correspond to the 99 and 10-minute GBT observations at 820~MHz, respectively. In Session 3, egress occurs halfway through the 99-minute observation, at $\phi_{\rm orb} = 0.50$. Sessions 34 and 41 correspond to Arecibo observations at 1430~MHz. Arecibo sessions at 327~MHz typically consist of several 20-minute observations made in succession. In some cases, the last observation is shorter if it was aborted at the end of assigned telescope time.  \label{fig_det}}
\end{center} 
\end{figure}

Many observations spanned eclipse ingress, egress, or showed
significant short-timescale emission variations, especially near
ingress (Fig.~\ref{fig_ingress}). We did not detect any radio emission in 
the orbital phase range $\phi_{\rm orb} = $~0.02--0.49, corresponding to the main 
eclipse when the pulsar is near superior conjunction with respect to the
companion. Sporadic ``mini-eclipses'' occur frequently 
at other orbital phases and their duration ranges from tens to hundreds of 
seconds. At two epochs we did not detect the pulsar even at orbital phases where it is normally visible, $\phi_{\rm orb} \sim$~0.6--0.8 (MJD 57343 and 57345; sessions 36 and 37 in Fig.~\ref{fig_det}), indicating that the pulsed radio emission may be obscured by dense ionized gas for extended periods of time. 

\subsection{Timing Solution}

\begin{figure}
\begin{center}
\includegraphics[width=0.75\textwidth, angle=-90]{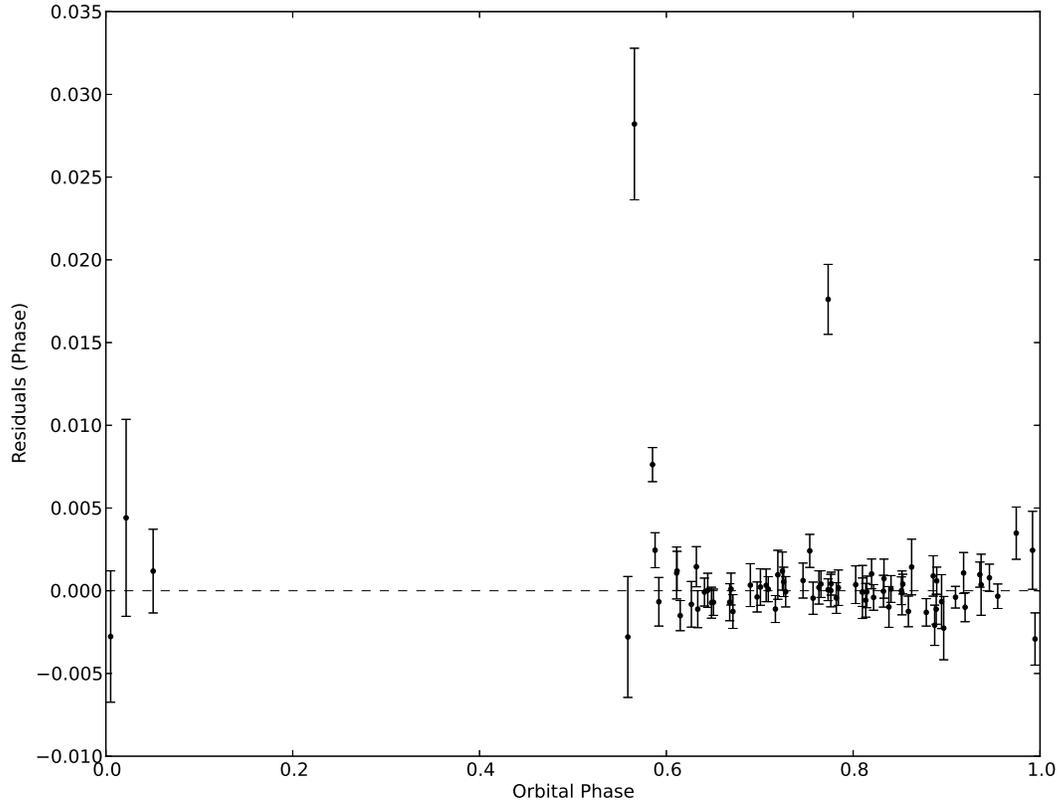}
\caption{Timing residuals vs.\ orbital phase, showing an eclipse over
  approximately half the orbit. All points shown are from observations
  with the Arecibo telescope at 327~MHz. Typical observing time per TOA 
  is 20 minutes.
\label{fig_res}}
\end{center}
\end{figure}

In order to derive a phase-connected timing solution that accounts for
every rotation of the pulsar between observations, we extracted
times-of-arrival (TOAs) from folded 327~MHz data using a pulse profile
template consisting of four Gaussian components, two each for the main
pulse and interpulse. 
We used the \textsc{Tempo} software package\footnote{http://tempo.sourceforge.net} and the BTX binary model
(\citealt{BT76}; D.~J.~Nice, unpublished) to perform least-squares fitting
of various pulsar parameters by minimizing the square of the
differences between expected and measured TOAs, after excluding three points with large, apparently dispersion-induced delays. We scale the TOA uncertainties by a factor of 1.47 in order to bring the reduced $\chi^2$ of the fit to unity.
The resulting best-fit parameters and other derived pulsar properties
are listed in Table \ref{tab_pulsars}. Figure~\ref{fig_res} shows the timing residuals for all 327~MHz observations. There are no significant trends
in the remaining timing residuals, indicating that at the current timing
precision and for the current data set, the ephemeris presented in
Table \ref{tab_pulsars} describes the TOAs. The most unusual timing property
of J1048+2339 is its orbital period variation necessitating the
fitting of five orbital period derivatives in order to obtain a timing
model that fits the data well over the full time span of radio
observations ($\sim 2$ years).

The relatively short span of the
solution and the large number of orbital period derivatives required
to model the data mean that there are significant correlations between
some of the timing parameters (e.g. the astrometric and orbital
parameters). As a result, the \textsc{Tempo} uncertainties of the celestial coordinates are significantly underestimated (see Section~\ref{sec:energetics}).
 
Two TOAs near eclipse egress (orbital phase $\sim 0.55$) show an extra delay indicating that pulses are passing through the outer layers of the companion. One TOA near orbital phase $\sim 0.75$ also shows an extra delay. Such
features are common in systems where gas turbulence and outflows may
introduce short-lived dispersive delays in the radio emission at
various orbital phases. We note that this feature is present in only
one of several observations spanning that part of the orbit.
We calculate that the extra delays observed for these three TOAs correspond to a DM excess of 0.008, 0.003, and 0.005~pc~cm$^{-3}$. Assuming that the DM excess is incurred over a linear distance equivalent to the projected semimajor axis of the orbit (or the radius of the companion, which is of the same order of magnitude), we obtain an intra-binary electron density on the order of $10^5-10^6$~cm$^{-3}$. In comparison, \cite{nt92} find an intra-binary electron density of $\sim 10^7$~cm$^{-3}$ for the eclipsing MSP B1744$-$24A, and 
\cite{Ransom04} find a similar value for PSR~J2140$-$2310A.

\subsection{Polarimetry and Profile Evolution}

We performed three observations of PSR~J1048+2339 using the Arecibo L-wide receiver and the PUPPI backend with a bandwidth of 700~MHz centered on 1430~MHz. One of the observations was in search mode, for a duration of 15 minutes. The remaining two observations were in coherent dedispersion full-Stokes mode, for 40 minutes each. 

\begin{figure}
\begin{center}
\includegraphics[width=0.7\textwidth,angle=-90]{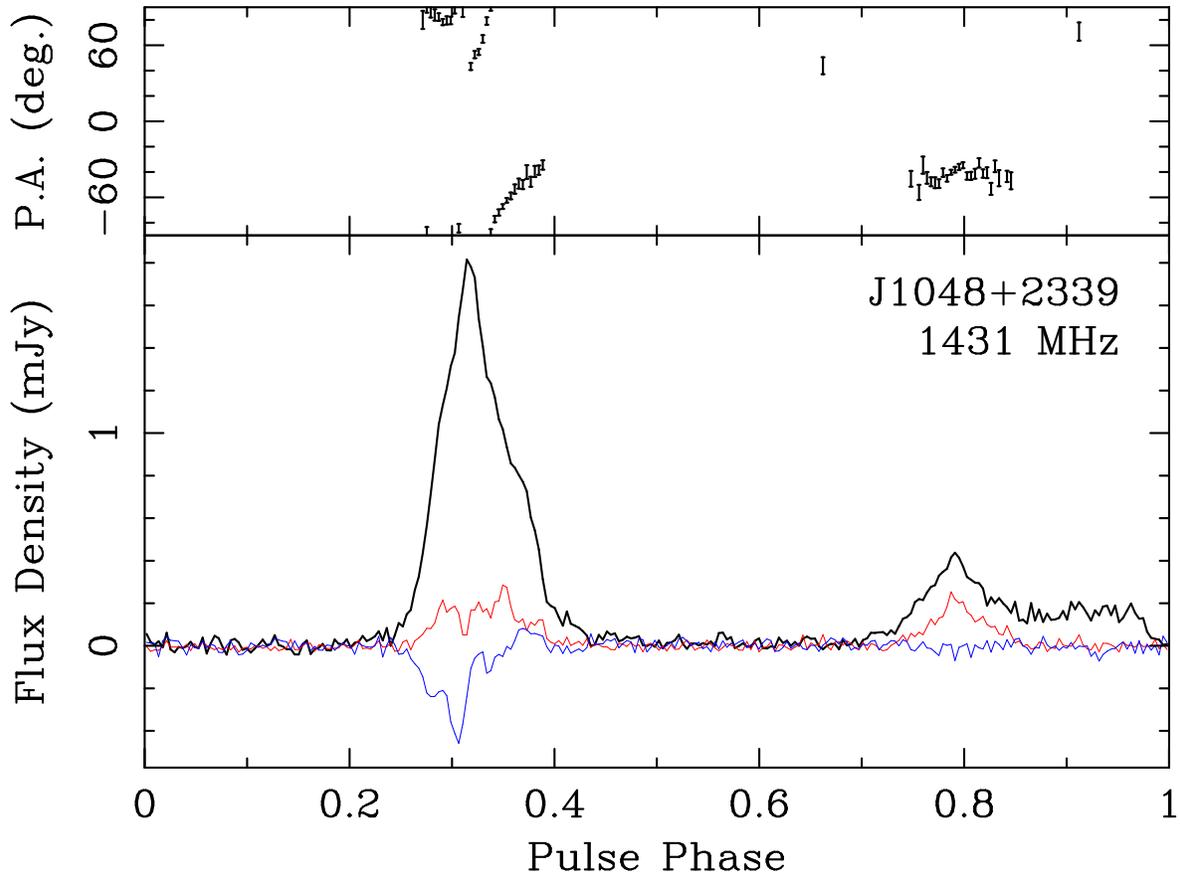}
\caption{Average polarimetric pulse profile for PSR~J1048+2339 from two
  40-minute Arecibo observations at 1430~MHz. In the bottom panel,
  total intensity is shown in black, while linear and circular
  polarization are shown in red and blue, respectively. In the top
  panel, the position angle of linear polarization is plotted for bins
  in the linear polarization profile with signal-to-noise ratio $>3$.
  \label{fig_pol}}
\end{center}
\end{figure}

The coherently dedispersed L-wide data were analyzed with PSRCHIVE
\citep{psrchive} and Figure~\ref{fig_pol} shows the calibrated 
full-Stokes pulse profile averaged from the two observations. 
The brightest component of the main pulse exhibits
significant circular polarization, while the interpulse is
linearly polarized. The pulse profile and linear polarization vector position 
angle are consistent between the two observations. The calibrated period-averaged flux densities are $S_{\rm 1430} = 0.076 \pm 0.02$~mJy and $0.199 \pm 0.04$~mJy, assuming a 20\% uncertainty. The estimated rotation measure is $-2.4 \pm 1.6$~rad~m$^{-2}$, consistent with zero. 

We computed the flux density from our
search-mode observations at other frequencies (where calibrated flux
density measurements were not available) using the radiometer equation
\citep{Handbook}. At 327~MHz, out of 80 data sets we excluded those
spanning ingress, egress, or containing ``mini-eclipses'' and calculated
a period-averaged flux density from each of the remaining 47 data sets.
The 327~MHz Arecibo receiver has gain $G = 11$~K/Jy\footnote{http://www.naic.edu/\~{}astro/RXstatus/rcvrtabz.shtml}. Nominal receiver temperature is 95~K, including spillover\footnote{http://www.naic.edu/\~{}phil/cal327/327Calib.html}. Scaling the 408~MHz sky temperature \citep{Haslam82} to 327~MHz yields $T_{\rm sky} = 33$~K.
Overall, $T_{\rm sys} = T_{\rm rec} + T_{\rm sky} = 128$~K. The range of flux density values we obtain is $S_{\rm 327} = $~0.5--5.3~mJy, with a mean of 1.6~mJy and a standard deviation of 1.1~mJy. 

We also analyzed two Robert C.\ Byrd Green Bank Telescope (GBT) observations
at 820 MHz from an earlier search of the {\em Fermi} source
3FGL~J1048.6+2338 where the pulsar was not detected in a blind
periodicity search, but a pulsed signal became evident after the data
were folded with our radio ephemeris. The observations used the GUPPI
backend with a bandwidth of 200~MHz. The 820~MHz receiver has $G =
2$~K/Jy and $T_{\rm sys} = 29$~K\footnote{http://wwwlocal.gb.nrao.edu/gbtprops/man/GBTpg/node14\_mn.html}. The observation times are 10 and 99 minutes, and we obtain period-averaged flux densities of $S_{\rm 820} = 0.62$ and 0.11~mJy, respectively.

The difference in flux densities between observations at each of the three observing frequencies indicates significant scintillation and/or variability. Because of this, and the fact that we have only two observations at 820~MHz and three at 1430~MHz, we refrain from calculating a spectral index, but it appears compatible
with that of a typical pulsar.

\begin{figure}
\begin{center}
\includegraphics[width=0.9\textwidth]{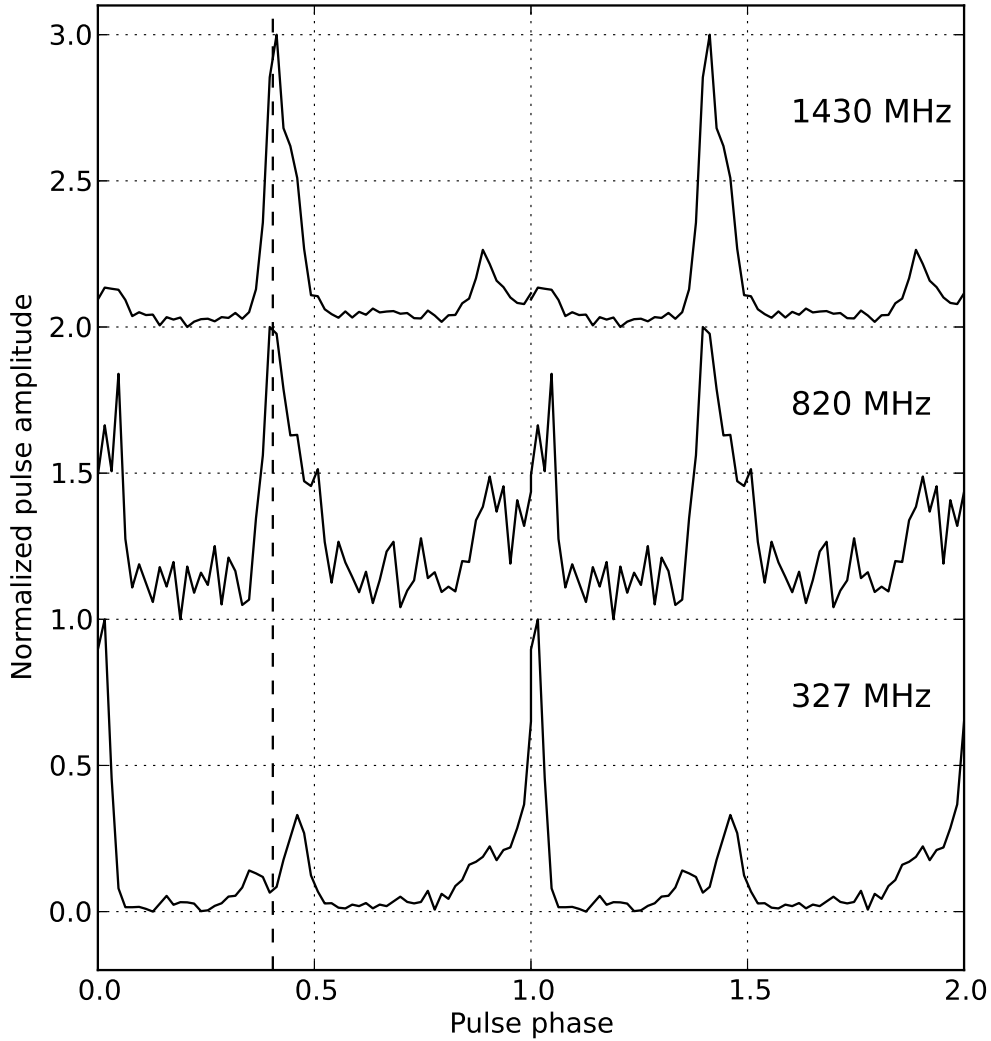}
\caption{Normalized averaged pulse profiles from a 40-minute
  coherently folded Arecibo observation at 1430~MHz, an archival
  10-minute search-mode GBT observation at 820~MHz, and a
  20-minute search-mode Arecibo observation at 327~MHz. The
  latter was selected for the absence of eclipse ingress, egress,
  short time scale emission variations, or TOA delays during the
  observation. The profiles are phase-aligned using the radio timing
  solution, and the dashed line provides a visual reference for the 
  intrinsic alignment of the peaks at different frequencies. Two complete turns 
  are shown.\label{fig_multifreq}}
\end{center}
\end{figure}

Figure~\ref{fig_multifreq} shows averaged pulse profiles at 327, 820,
and 1430~MHz aligned in phase using the radio ephemeris. There is
significant profile evolution with frequency: the pulse at phase $\sim
1.0$ is brighter at 327~MHz but becomes weaker than the interpulse as
the frequency increases. Within the interpulse at phase $\sim 0.4$ in Figure~\ref{fig_multifreq},
the trailing component is brighter at 327~MHz but the leading
component dominates at the higher frequencies.

\section{Optical Observations}\label{sec_optical}

\subsection{Catalina Real Time Transient Survey}
\label{sec:crts}

Using the radio timing position, we identified a 20th magnitude candidate optical counterpart to \psr\ in the Catalina Real-Time Transient Survey (CRTS, \citealt{dra09}) second data release. The Catalina ID number of the star is 1123054021805.

\begin{figure}
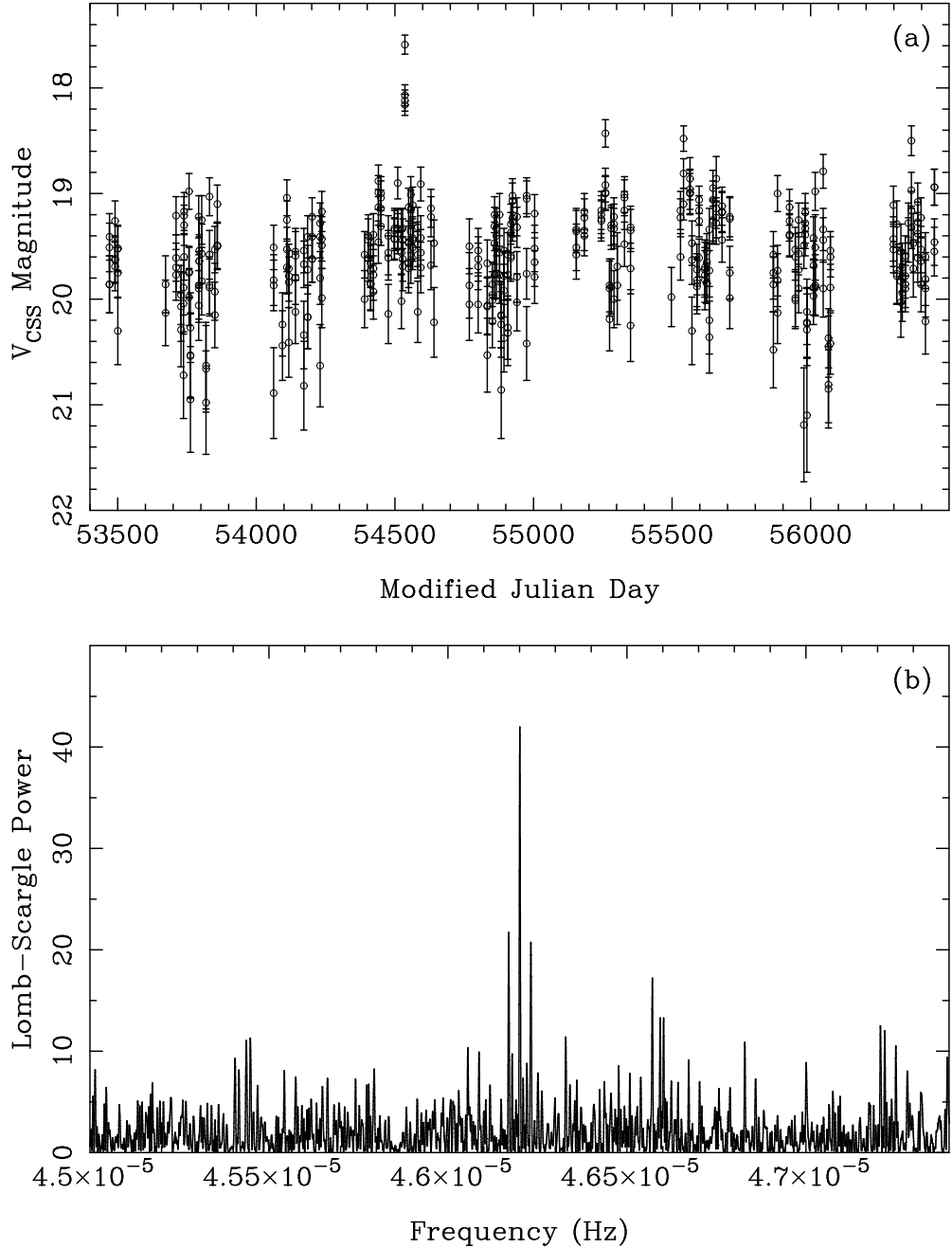

\centerline{
\includegraphics[angle=0,width=0.80\linewidth,clip=]{FIG6a.ps}
}
\vspace{0.2in}
\centerline{
\includegraphics[angle=0,width=0.80\linewidth,clip=]{FIG6b.ps}
}
\caption{
(a) Photometry of \psr\ from the CRTS, comprising 403
measurements on 119 nights spanning 2005 April 10 -- 2013 June 3.
(b) Lomb-Scargle periodogram of the above data excluding 25 outlier
points with $V_{\rm CSS}<19.0$.
The peak at $4.62007(3)\times10^{-5}$~Hz = 0.2505173(16)~d is consistent
with the radio pulsar orbital period.  Sideband frequencies are the 1-year
aliases.
}
\label{fig:periodogram}
\end{figure}

We extracted 8 years of CRTS data on this star,
consisting of 403 observations on
119 nights from 2005 April~10 to 2013 June~3
(Figure~\ref{fig:periodogram}a).  On a typical night, four unfiltered
30~s exposures of a field are taken within 30 minutes, and the
photometry is expressed as $V_{\rm CSS}$ magnitudes (discussed in
Section~\ref{sec:interpret}).  The $3\sigma$ limiting magnitude is
$V_{\rm CSS}\approx20.3$ as we infer from those points whose error
bars are $\pm0.3$~mag.

We first transformed the times of the data to barycentric
dynamical time for consistency with the radio pulsar ephemeris.
Magnitudes were converted to fluxes, which were normalized by
the variance of the data \citep{hor86}, and a Lomb-Scargle
periodogram \citep{sca82} was calculated around the radio
pulsar's orbital period.  In this analysis we
omitted high points that are clear outliers from the
mean light curve, because they add noise at all frequencies.
This was done crudely by excluding the 25 points with $V_{\rm CSS}<19.0$.
Although some may just be statistical fluctuations,
19 of them can be considered
confirmed outliers because there are at least two
on the same night. Examples of excluded points are
the four brightest in Figure~\ref{fig:periodogram}a,
with magnitudes 18.12--17.59 on MJD 54536 (2008 March 11).
They clearly represent some sort of flare, a non-stationary process.
This simple procedure may cause us to underestimate the normal
maximum flux, and it does not account for smaller flares that may
occur during fainter phases of the orbit; i.e., we do not iterate
the selection of outliers based on their phase in the folded
light curve.
 
A strong signal (Figure~\ref{fig:periodogram}b)
is detected at $0.2505173(16)$~d, consistent with the
more precise orbital period of the radio pulsar, $P_{\rm orb}=0.250519045(5)$~d.
The noise power in the periodogram is measured to be exponentially distributed with
a mean of 2.06 (not 1 as expected for white noise, due to the
closely correlated times of the nightly points), so the peak power
of 42.0 has a single-trial chance probability of
$e^{-42.0/2.06}=1.4\times10^{-9}$.

\begin{figure}
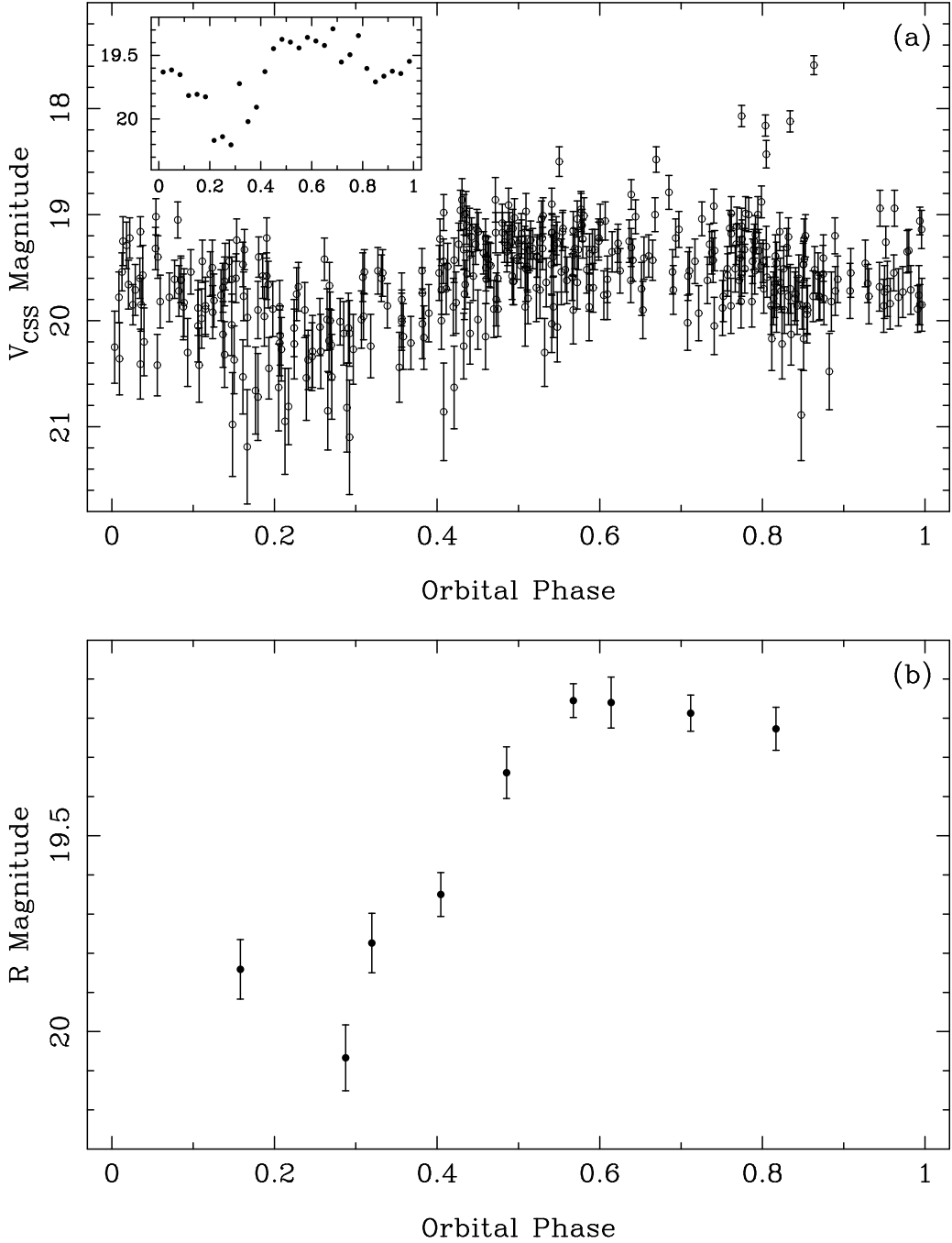

\centerline{
\includegraphics[angle=0,width=0.83\linewidth,clip=]{FIG7a.ps}
}
\vspace{0.2in}
\centerline{
\includegraphics[angle=0,width=0.83\linewidth,clip=]{FIG7b.ps}
}
\caption{
Folded light curves of \psr\ using the (constant) orbital period and phase
from the radio pulsar ephemeris.  Phase 0 corresponds to the ascending
node of the pulsar.  (a) CRTS data from Figure~\ref{fig:periodogram}a.
The inset is the binned light curve, constructed after excluding points
with $V_{\rm CSS}<19.0$.  (b) PTF data.  All points were obtained over
2012 March 23--27.
}
\label{fig:fold}
\end{figure}

Since the optical search found a period consistent with the radio pulsar's
orbital period, we folded the data on the precise radio ephemeris.
More precisely, we used the orbital period and time of ascending node,
but not the period derivative(s), as the latter are known to wander on
long timescales in such systems
and can become either positive or negative, as seen in
\citet{arz94} and \cite{ple15}.  For the purpose of displaying the
folded data (Figure~\ref{fig:fold}a), we restored
the high points previously omitted as flares in the
periodogram analysis.
Finally, we calculated a mean light curve
by averaging the points (by flux, but not weighted by error bars)
in 30 bins in orbital phase while
excluding the flaring points.  The mean light curve
is displayed in the inset of Figure~\ref{fig:fold}a.

\subsection{Palomar Transient Factory}
\label{sec:pft}

We also extracted observations of \psr\ from the Palomar Transient
Factory (PTF, \citealt{law09,rau09}).  Only nine $R$-filter measurements
were made over 2012 March 23--27, but they fortuitously cover a large
fraction of the orbit, with smaller errors than CRTS.
As shown in Figure~\ref{fig:fold}b, the shape of the modulation is very
much like the mean CRTS light curve, and has the same total
amplitude of $\approx0.9$~mag.

\subsection{Sloan Digital Sky Survey}
\label{sec:sdss}

The Sloan Digital Sky Survey (SDSS, \citealt{SDSS}) photometric catalog lists
the optical counterpart of PSR~J1048+2339 at position
(J2000.0) R.A.$=10^{\rm h}\,48^{\rm m}\,43^{\rm s}\!.4270(5)$,
decl.$=+23\degree\,39\amin\,53\asec\!.503(7)$.  
The observation occurred on 2005 February 4 (MJD 53405.39)
at phase 0.96 of the binary orbit.
The SDSS PSF magnitudes are $u=23.45\pm0.46,
g=20.819\pm0.040, r=19.646\pm0.024, i=19.029\pm0.022, z=18.733\pm0.034$.
We used the equations of Lupton (2005)\footnote{https://www.sdss3.org/dr8/algorithms/sdssUBVRITransform.php\#Lupton2005}
to transform these to the Johnson system, giving $V=20.14, B-V=1.27, V-R=0.82$.
This is useful, as the CRTS unfiltered magnitude $V_{\rm CSS}$ is not
a good approximation to Johnson $V$ for red stars.
Knowing the color, a more precise calibration of CRTS
is given as $V=V_{\rm CSS}+0.91(V-R)^2+0.04$ \citep{dra13}.
At $\phi_{\rm orb}=0.96$, $V_{\rm CSS}\approx19.6$ from the
binned light curve in Figure~\ref{fig:fold}a, therefore
$V=V_{\rm CSS}+0.65=20.25$, in excellent agreement with $V=20.14$
from the SDSS.  The total range of the light curve,
neglecting any color changes, is then $20.0 < V < 20.9$.
In principle, one could get some information about the $V-R$
color variation around the orbit by comparing the CRTS and PTF
light curves.  But the sparseness of the PTF data, and the poor
sensitivity of the CRTS data around the minimum of the light curve,
do not enable such a variation to be measured robustly.


\subsection{Pan-STARRS}
\label{sec:panstarrs}

PSR~J1048+2339 was also detected by the Pan-STARRS PS1 survey \citep{Chambers06, Kaiser06}. The Pan-STARRS survey has been photometrically calibrated using a two-stage approach.  In the first stage, photometrically stable nights are identified and the system zero points and a modest number of other parameters are determined based on the relative photometry overlaps, resulting in systematic uncertainties of (\gps, \rps, \ips, \zps, \yps) = (8.0, 7.0, 9.0, 10.7, 12.4) mmag \citep{Schlafly12}.  In the second stage, the resulting zero points are used to tie the remainder of the data to this internal system \citep{Magnier13}, with the overall photometric system tied to the AB system via spectrophotometric standards \citep{Tonry12}.

There have been to date three image processing pipeline versions, designated PV1, PV2, and PV3. PV2 is currently the most reliable astrometrically, but yields fewer total detections than PV3 because it covers fewer epochs. The shorter wavelengths give smaller uncertainties in position. Using 16 detections in the \gps\ and \rps\ filters (mean epoch MJD 55674) and assuming a 50~mas error radius to allow for unknown systematics, we find (J2000.0) R.A.$=10^{\rm h}\,48^{\rm m}\,43^{\rm s}\!.429(3)$, decl.$=+23\degree\,39\amin\,53\asec\!.49(5)$.

\begin{figure}
\centerline{
\includegraphics[angle=0,width=0.44\linewidth,clip=]{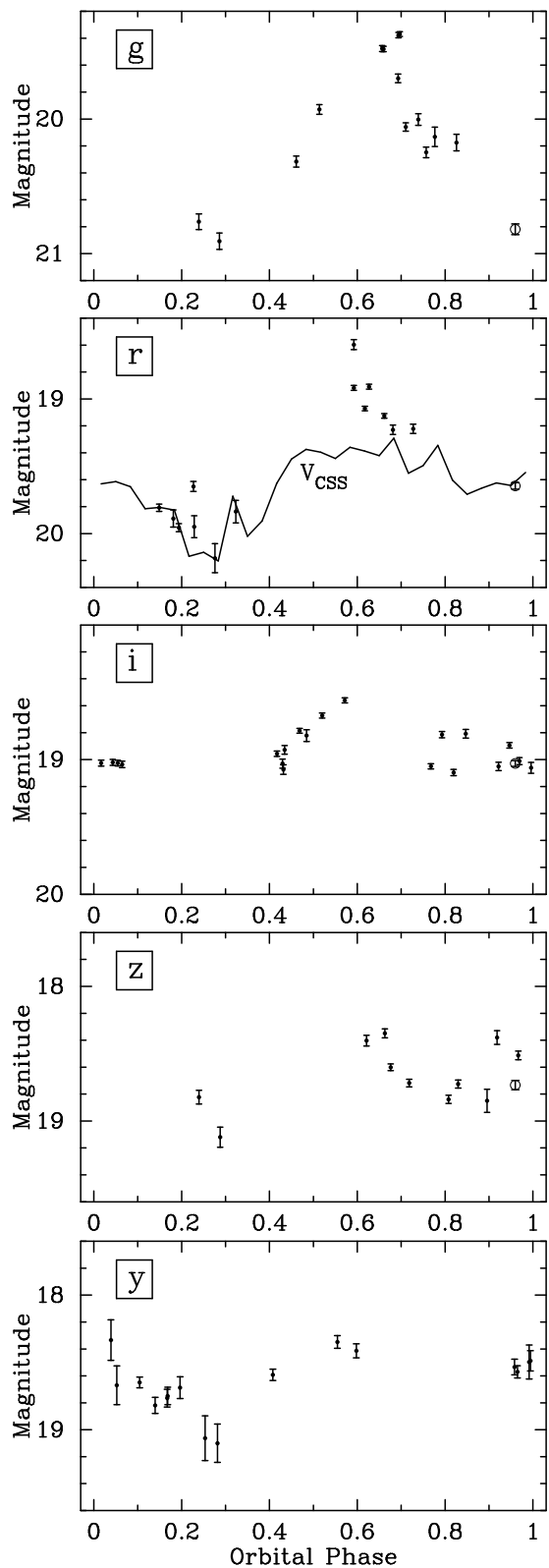}
}
\caption{Folded light curves of PSR~J1048+2339 in the Pan-STARRS survey. The vertical axis spans exactly 2 magnitudes in each panel to illustrate the relative amplitude of variability in the five bands.  The solid line is the mean unfiltered ($V_{\rm CSS}$) light curve from the inset of Figure~\ref{fig:fold}a.  The open circles at $\phi_{\rm orb}=0.96$ are the SDSS $g, r, i, z$ points.}
\label{fig_grizy}
\end{figure}

PV3 is preferred for photometry, being better calibrated and more complete.  For PSR~J1048+2339, PV3 yields 75 useful detections (by a point-spread function goodness-of-fit criterion), typically obtained on 6--8 nights per filter, and spanning MJD~$55242 - 57043$ (2010 February 15 -- 2015 January 21).  Figure~\ref{fig_grizy} shows the light curves of PSR~J1048+2339 in all five bands, with orbital phase calculated using the same method as for the CRTS and PTF data (see Section~\ref{sec:crts}).  Phase coverage is non-uniform in all filters. Nevertheless, it is evident that the amplitude of modulation is a decreasing function of wavelength; \gps\ and \rps\ span more than 1.0 mag, while the other bands are less variable.  This is expected for a heating effect, where brighter regions are hotter.  In all filters, the minimum occurs at $\phi_{\rm orb}\approx0.25$ and the maximum at $\phi_{\rm orb}\approx0.6$. On the \rps\ filter data in Figure~\ref{fig_grizy} we superpose the mean unfiltered CRTS light curve from the inset of Figure~\ref{fig:fold}, without any renormalization.  These two light curves bear some resemblance, although both \gps\ and \rps\ have considerably higher amplitude of modulation than CRTS. The small number but high accuracy of the Pan-STARRS points leaves an impression that intrinsic variation may be frequent, while it is smoothed out in the CRTS mean light curve that comprises a much larger number of less precise measurements.

\section{Gamma-ray Counterpart}\label{sec_gamma}

We performed a spectral analysis of the gamma-ray source
3FGL~J1048.6+2338 to characterize its spectrum and perform an
optimized pulsation search using the weighted H-test
\citep{Kerr11}. We used a subset of the recently released Pass 8 LAT data set
starting from 2013 August 4 (the beginning of our phase-connected radio timing solution) and extending through 2015 July 15 (MET 397267203 -- 458611204).
We selected SOURCE class events (\texttt{evclass = 128
  and evtype = 3}) only during the intervals of good science data
(\texttt{DATA\_QUAL=1 and LAT\_CONFIG=1}) and restricted our events to
those with a zenith angle less than 90$\degr$ and energies from 100~MeV to 100~GeV.  We selected events from a 15$\degr$ radius around the source and
performed a binned likelihood analysis over a $20\degr \times 20\degr$
region with 0.1$\degr$ pixels. The initial model we used was based on
the 3FGL catalog \citep{3FGL}. We modified the spectral model for the
target source to be an exponentially cutoff power law of the form 
\be
\frac{dN}{dE} = N_0 \frac{E}{E_0}^{-\Gamma}
\exp\left(-\frac{E}{E_c}\right).
\ee

\begin{figure}
\begin{center}
\includegraphics[width=0.9\textwidth]{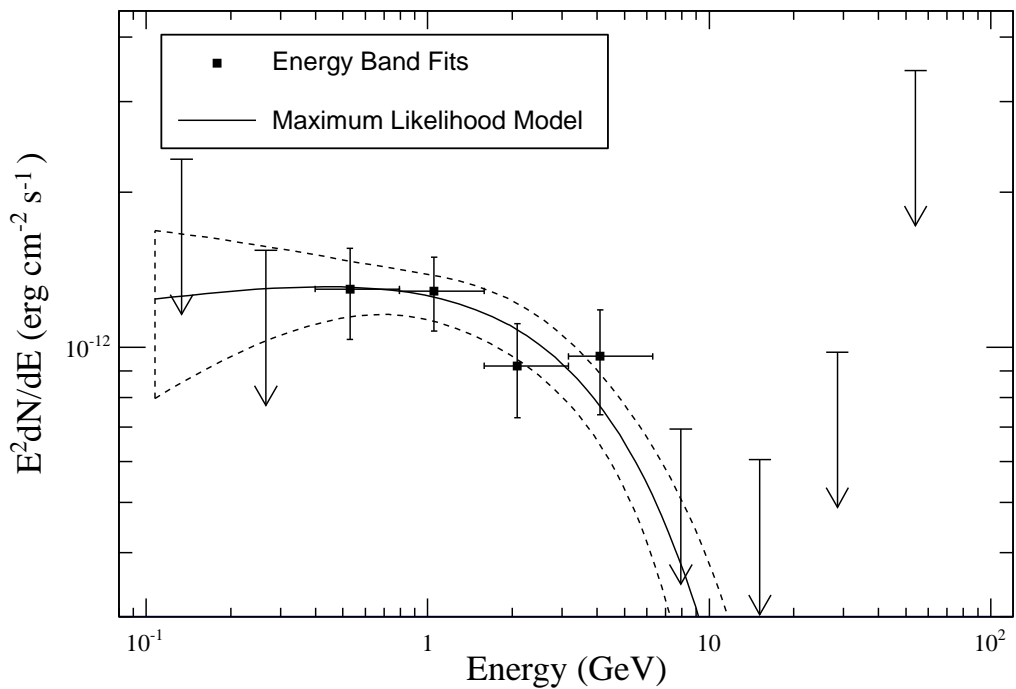}
\caption{The maximum likelihood fit of an exponentially cutoff power
  law to the gamma-ray spectrum of PSR~J1048+2339 is shown as a solid line
  and gives a photon spectral index of $1.9 \pm 0.2$ and a cutoff energy of
  $5.0 \pm 2.3$~GeV. The dashed curves represent the fit uncertainty. The 
  points and upper limits are the fitted flux densities in 10 energy
  bands. 
\label{fig_gspec2}}
\end{center}
\end{figure}

To perform the maximum likelihood fit, we used the
\texttt{P8R2\_SOURCE\_V6} instrument response functions with the 
\textit{Fermi} Science Tools version 10-00-05 and the \texttt{NewMinuit}
fitting function. In the initial fit, we held all parameters fixed at the
3FGL catalog values except for the spectral parameters for the target
source, and the normalization for sources within 6$\degr$ of the
target or flagged in the 3FGL catalog as being variable. With the
exponentially cutoff power-law (ECPL) model, the source is detected
with a test statistic (TS) of 228, corresponding to 15 $\sigma$ \citep{mattox96}.
Integrating the spectral model over the range 100 MeV to 100 GeV yields a
 photon flux of $7.8(2) \times 10^{-9}$~ph~cm$^{-2}$~s$^{-1}$ and an energy flux
 of $5.0(6) \times 10^{-12}$~erg~cm$^{-2}$~s$^{-1}$. The photon spectral index is $1.9 \pm 0.2$
and the cutoff energy is $5.0 \pm 2.3$~GeV. Figure~\ref{fig_gspec2}
shows the spectral fit. When we fit with a simple power law and integrate over
the same energy range, we find a photon flux of $1.04(15) \times 10^{-8}$~ph~cm$^{-2}$~s$^{-1}$, an energy flux of $6.4(6) \times 10^{-12}$~erg~cm$^{-2}$~s$^{-1}$, and a spectral index of $2.29 \pm 0.07$. The ECPL model is weakly preferred with the cutoff detected at a significance of only 1.3 $\sigma$.

For the gamma-ray timing analysis, we selected photons from a region
of radius 2$\degr$ around the pulsar and assigned weights based on the
best-fit spectral model. We compute a pulse phase for each selected
LAT photon using the \texttt{fermi} plugin for \textsc{Tempo2} \citep{Ray11} and the
best-fit radio timing model.  The resulting weighted H-test value was
18.8, corresponding to a significance of 3.45$\sigma$, not enough for a
high-confidence claim of a pulsed detection. It is not feasible to extend
the span of gamma-ray timing analysis earlier than the start of our radio
timing solution due to the significant orbital period variations that are
the hallmark of PSR~J1048+2339. However, the H-test value improves
as the radio timing solution and correspondingly the gamma-ray timing span
are extended forward with the accumulation of more radio observations. This
suggests that a confident detection of gamma-ray pulsations is likely with
$1-2$ more years of radio timing data and an improved timing model that 
includes proper motion and possibly higher-order orbital period derivatives. 

\section{No X-ray Detection}

\begin{figure}
\begin{center}
\includegraphics[width=0.9\textwidth]{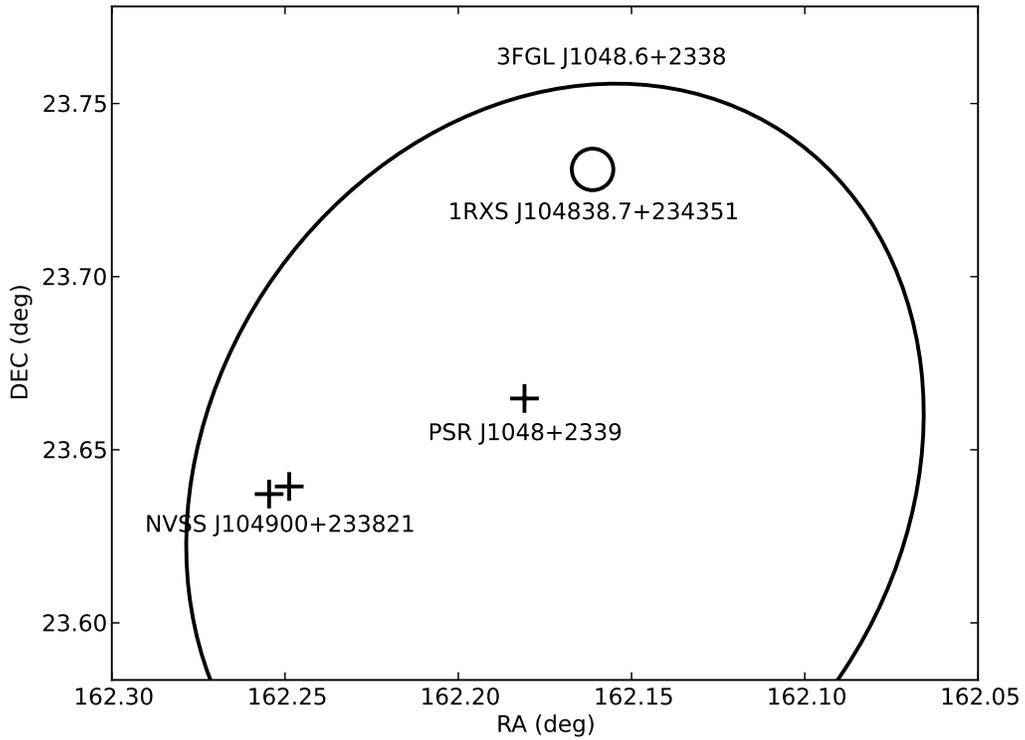}
\caption{Relative positions of the 3FGL source J1048.6+2338 error
  ellipse denoting 95\% confidence, PSR~J1048+2339, and
  the two lobes of the AGN NVSS~J104900+233821. The position
  uncertainty of PSR~J1048+2339 from timing is much smaller than the
  marker size. Also shown is the error circle of the X-ray source
  1RXS~J104838.7+234351. \label{fig_sources}}
\end{center}
\end{figure}

A single {\em Swift} XRT observation exists of the field in Figure~\ref{fig_sources}, taken on 2011 January 19 in Photon Counting mode for an exposure of 1.2 ks (ObsID 00041540001).  
The mid-time of the exposure falls at $\phi_{\rm orb}=0.62$, which is near the maximum in the optical light curves. 
Only three counts were detected inside a circular radius 47\asec\ around the position of PSR~J1048+2339 which is consistent with a non-detection of any X-ray source.
We derive an absorbed 0.5--3.0~keV 3-$\sigma$ upper limit of 
$1.3 \times 10^{-13}$~erg~s$^{-1}$~cm$^{-2}$ for a 0.2~keV blackbody neutron 
star surface with the estimated Galactic hydrogen column density for the pulsar position, $N_{\rm H} = 2.3 \times 10^{20}$~cm$^{-2}$ \citep{Kalberla05}. For the AGN NVSS~J104900+233821 we derive an upper limit of $1.7 \times 10^{-13}$~erg~s$^{-1}$~cm$^{-2}$ for an assumed X-ray power-law index of 1.8 and the same  Galactic hydrogen column density. 

The X-ray source 1RXS~J104838.7+234351, also shown
in Figure~\ref{fig_sources}, is within the error ellipse of the {\em Fermi} 
source 3FGL~J1048.6+2338. Since the position uncertainties for both the X-ray
source and PSR~J1048+2339 are much smaller than the distance between
the two objects, an association can be ruled out.

\section{Discussion}

\subsection{Association with the Gamma-ray Source}

The sources for our radio search were selected from a preliminary
version of the {\em Fermi} 3FGL catalog \citep{Cromartie15}. 
In the final version of this
catalog \citep{3FGL}, the source 3FGL~J1048.6+2338 is
listed as associated with the BL Lac active galactic nucleus
NVSS~J104900+233821. However, the association was made by a
pipeline based on positional proximity and the radio flux 
density distribution of NVSS sources. While most pulsars exhibit gamma-ray spectra that can be fitted by a power law with an exponential cutoff, for weaker gamma-ray
sources the spectrum shape may be ambiguous and be equally well-fitted with a power law, which is typical of AGNs. 
The discovery of PSR~J1048+2339 well
within the {\em Fermi} source error ellipse (Figure~\ref{fig_sources})
means that the AGN association is likely spurious. A secure detection of
gamma-ray pulsations from this pulsar will confirm the identification with
3FGL~J1048.6+2338 and allow a search for emission from the NVSS source
by gating the data to include only the off-pulse phase of the pulsar.

\subsection{Position, Proper Motion, and Pulsar Energetics}
\label{sec:energetics}

The SDSS and Pan-STARRS optical positions are 
(J2000.0) R.A.$=10^{\rm h}\,48^{\rm m}\,43^{\rm s}\!.4270(5)$, decl.$=+23\degree\,39\amin\,53\asec\!.503(7)$
at epoch 2005 and
(J2000.0) R.A.$=10^{\rm h}\,48^{\rm m}\,43^{\rm s}\!.429(3)$, decl.$=+23\degree\,39\amin\,53\asec\!.49(5)$
at epoch 2011, respectively. While they are consistent within their
respective uncertainties, the radio timing position in
Table~\ref{tab_pulsars}, with end figures $43^{\rm s}\!.4168$(1) and $53\asec\!.403(4)$,
is offset from them by $\approx200$~mas. 

The PPMXL catalog \citep{PPMXL} includes the optical counterpart of
PSR J1048+2339, with components of proper motion $\mu_\alpha{\rm cos}\,\delta = -7.8 \pm
6.2$~mas~yr$^{-1}$, $\mu_\delta = -9.7 \pm 6.2$~mas~yr$^{-1}$.
The APOP catalog of proper motions outside of the
Galactic plane \citep{APOP} also has an entry for this object, with
$\mu_\alpha{\rm cos}\,\delta = -18.7 \pm 3.0$~mas~yr$^{-1}$, $\mu_\delta = -9.4 \pm
0.7$~mas~yr$^{-1}$.  The discrepancies between these results give some idea
of the systematic errors involved, since they both derive from the same plate
material, the Palomar Observatory Sky Survey.


The pulsar position, proper motion, and orbital
period derivatives are highly covariant in the timing analysis. 
Despite having two years of
radio timing data on PSR~J1048+2339, we are still not able to
confidently fit for proper motion along with the five orbital period
derivatives listed in Table~\ref{tab_pulsars}. This means that any
actual proper motion is subsumed in the fits for $f_{\rm
  orb}^{\rm(n)}$ and position. The timing solution in
Table~\ref{tab_pulsars} uses the APOP proper motion as a fixed parameter and 
results in the smallest rms residual. However, we are
also able to obtain phase-connected timing solutions using several
alternative sets of free parameters: (1) position, proper motion, and
$f_{\rm orb}^{\rm(1)}$, (2) position, $f_{\rm orb}^{\rm(1)}$, and
$f_{\rm orb}^{\rm(2)}$, and (3) position, $f_{\rm orb}^{\rm(1)}$,
$f_{\rm orb}^{\rm(2)}$, and $f_{\rm orb}^{\rm(3)}$. Based on the
spread of position fits between these timing solutions, we conclude
that the position in the timing solution listed in
Table~\ref{tab_pulsars} is consistent with the SDSS and Pan-STARRS
optical positions within a remaining systematic error of $\sim
250$~mas.

The present-epoch error ellipse for the PPMXL position, taking into account
the proper motion uncertainty, encompasses best-fit positions from all
of our attempted timing solutions described above. The present-epoch
error ellipse for the APOP position is within our systematic error of
$\sim 250$~mas from the best-fit positions of all timing
solutions.  When it becomes possible to fit confidently for proper motion in
our timing solution, the resulting position will very likely converge
to that of the optical counterpart.

The observed rotation period derivative $\dot{P}$ includes
a kinematic component $\dot{P}_{\rm k} = P \mu^2D/c$ \citep{Shklovskii70},
where $\mu$ is the total proper motion.
At the NE2001-estimated distance of 0.7~kpc, we calculate
$\dot{P}_{\rm k} = 1.2 \times 10^{-21}$~s~s$^{-1}$ using PPMXL and
$\dot{P}_{\rm k} = 3.5 \times 10^{-21}$~s~s$^{-1}$ using APOP total proper
motions, which contribute, respectively, 4\% or 12\% of the observed
$\dot{P} = 3.0 \times 10^{-20}$~s~s$^{-1}$.
Given the evident systematic errors, we assume only an upper
limit on proper motion of $\mu<25$~mas~yr$^{-1}$, corresponding to
a tangential velocity $v_{\perp}<83$~km~s$^{-1}$ at $D=0.7$~kpc.
This would imply a $<17$\% reduction to the apparent pulsar spin-down
luminosity of $\dot{E} = 1.2 \times 10^{34}$~erg~s$^{-1}$ ($ \dot{E} \propto
\dot{P} P^{-3}$).  

The gamma-ray luminosity can be computed following \citet{2PC}, equation 15. Assuming $f_\Omega = 1$, and using the DM distance of 0.7 kpc, the gamma-ray luminosity is $L_\gamma = 2.9 \times 10^{32}$ erg s$^{-1}$. This yields an efficiency $\eta_\gamma = L_\gamma/\dot{E}$ of 2.4\%. This is consistent with, but on the low end of the distribution of, efficiencies observed for LAT-detected MSPs \citep{2PC}. Any kinematic corrections to the spin-down will serve to increase the efficiency.

\subsection{Interpreting the Optical Modulation}
\label{sec:interpret}

The most consistent feature of the optical light curves
is the minimum at $\phi_{\rm orb}\approx0.25$.
This accords with the expectation for pulsar wind
heating of the companion star, as phase 0.25 corresponds to inferior
conjunction of the companion, when we are viewing its cold side.
The consistent alignment also implies that a constant
radio period serves well to extrapolate the orbital phase over the
previous 9~yr.   On the other hand, the light curves do not have
the symmetric heating maximum around $\phi_{\rm orb}=0.75$ expected in this scenario.
Rather, the flux is higher at $\phi_{\rm orb}=0.6$ and decreases toward $\phi_{\rm orb}=1.0$.
This resembles a similar sloping feature in the optical and X-ray
light curves of the transitioning millisecond pulsar PSR J1023+0038 when
in its radio-pulsar state \citep{wou04,bog11}.
It could indicate that the heating
is asymmetric because it does not come directly from the pulsar, but is
mediated by an intrabinary shock that has a skewed shape near the companion.
Alternatively, the shape of the light curve may be affected by
the magnetic field of the companion directly channeling the
pulsar wind to localized regions on its surface \citep{tan14,li14}.

Spectral hardening around $\phi_{\rm orb}=0.6$ seen in the Pan-STARRS data is
generally consistent with heating due to irradiation from the pulsar.  The amplitude
of modulation is about 0.8 magnitudes in \ips, \zps, and \yps, but it increases
to $1.2-1.4$ magnitudes in \gps\ and \rps.  The departure from the expected
symmetry around $\phi_{\rm orb}=0.75$ in the Pan-STARRS data is even more
pronounced than in CRTS, which, because of the smaller number of Pan-STARRS
points, may indicate that stochastic variability is prevalent.
Some measurements in the same filter differ by as much as
$\sim0.4$~magnitudes at the same orbital phase.

There may be some significance to the fact that the peak in the
apparent heating effect is displaced toward orbital phases earlier than 0.75 in both CRTS and
Pan-STARSS.  An earlier peak implies that the trailing side of the companion, seen while
it is moving away from the observer, is hotter than the leading side. 
This may indicate where the intrabinary shock is closest to the
surface of the companion and illuminating it with the greatest
geometrical covering factor.  In his prediction for ``hidden''
millisecond pulsars of type II, \citet{tav91} argued that the wind ablated from
the companion would be displaced in the direction of orbital motion
(leading) due to the Coriolis force.  This would force the shock
away from the companion in the leading direction, possibly
leaving the more efficient heating to the trailing side.

Any more detailed interpretation of the optical modulation is hampered by the
evident intrinsic variability in the CRTS and Pan-STARRS data on timescales
that are not yet fully explored.  However, it would be interesting with new
data to relate well-sampled optical light curves to the orbital phase
distribution of the radio eclipses, to see if both can be explained
by the geometry of the ablated wind.

In addition to heating, there may be some modulation due to our changing
view of the tidally distorted surface of the companion, nearly filling 
its Roche-lobe.  Such ellipsoidal modulation produces maxima of equal
height at $\phi_{\rm orb}=0$ and 0.5, and unequal minima at $\phi_{\rm orb}=0.25$ and 0.75,
the latter one being lower.  This effect dominates the light
curve of the 5~hr period redback PSR J1628$-$3205 \citep{li14}
and others, e.g., \citet{bel13,hui15}.  However, the maximum
ellipsoidal amplitude is only expected to be $\sim0.4$~mag,
while the mean light curve of \psr\ varies by $\gtrsim1.0$~mag.  Also, the
two local minima in the mean CRTS light curve are not separated by 0.5 in phase,
but more like 0.6.  It is possible that ellipsoidal
modulation is a small contributor even though it cannot
explain the gross features of the light curve of \psr.

\subsection{The Companion Star}
\label{sec:companion}

We would like to know the intrinsic colors of the companion star
to estimate its spectral type and radius.   While the Pan-STARRS
photometry represents the best broad-band spectral information
currently available on this object, its sparse observations,
together with the substantial variability of the source,
both orbital and random, prevents a reliable color from being 
obtained from those exposures, which were taken more than 16 minutes
apart.  In contrast, the single-epoch SDSS photometry was acquired
within 5 minutes, which is why we used the magnitudes transformed
from the SDSS to calibrate the CRTS light curve
approximately (see  Section~\ref{sec:sdss}). 
The conclusions that follow from this are tentative,
and must be tested with detailed temperature measurements and
modeling of the orbital light curve. 

Assuming that we are viewing the orbit nearly edge-on ($i=90^{\circ}$),
$V\approx20.9$ at $\phi_{\rm orb}=0.25$ represents the unheated side of
the companion. At the DM distance of $D=0.7$~kpc, the interstellar extinction
measured from Pan-STARRS and 2MASS photometry is $A_V=0.05$ \citep{gre15},
so the extinction-corrected apparent magnitude is $V=20.85$
and the absolute magnitude is $M_V\approx +11.6$, suggesting a main-sequence
star of spectral type M4, which has a radius of $R_c=0.35\,R_{\odot}$
and an effective temperature of $T_{\rm eff}\approx3350$~K.
(If $i<90^{\circ}$, the phase of minimum light represents a mixture
of spectral types both hotter and cooler than M4.)  But the SDSS
colors at $\phi_{\rm orb}=0.96$, near quadrature, correspond to an earlier
spectral type, around K6.  Therefore, we can assume that the
heated side of the companion is, on average, hotter than a K6 star,
or $T_h>4200~K$. 

Given the pulsar's apparent spin-down luminosity,
$\dot E\approx1\times10^{34}$ erg~s$^{-1}$ corrected
for proper motion, we can estimate the
efficiency of pulsar heating during the bright phase
of the optical light curve.  Assuming an average $T_h=4400$~K
for the heated half of the star, its extra luminosity
at phase 0.75 can be expressed in terms of the heating efficiency
$\eta$ of an assumed isotropic pulsar wind as
$\eta\dot E R_c^2/4 a^2=2\pi R_c^2\sigma (T_h^4-T_{\rm eff}^4)$,
where $a$ is the orbital separation.  (Other authors use a factor
of $4\pi$ here instead of our $2\pi$.) For an assumed NS mass of
$1.4\,M_{\odot}$ and an orbital inclination of $i=90^{\circ}$,
$M_c=0.31\,M_{\odot}$ and $a=2.0\,R_{\odot}$. Therefore,
pulsar heating is only marginally able to power the observed
peak luminosity, because the needed efficiency $\eta$ is $\approx0.69$.

There are a number of theoretical reasons that $\eta$
should not exceed 0.5, as summarized by \citet{li14}.
However, \citet{bre13} estimated an efficiency in the range
$0.1\le\eta\le0.6$ for the four systems they modeled,
while \citet{sch14} found $\eta>1$ for the black
widow PSR J1810+1744.  \citet{rom15b} modeled the redback
PSR J2215+5135 with $\eta=0.68$, and \citet{rom15a} found
$\eta>1$ for the black widow PSR J1311$-$3430.  So more complex
physical effects are clearly having an important influence.
One way to reduce the inferred efficiency is to assume that
the pulsar wind is beamed in the orbital plane in this system,
or channeled by the intrinsic magnetic field of the companion.

With an estimated $R_c=0.35\,R_{\odot}$ from its optical properties,
the companion appears to significantly underfill its Roche lobe,
which has radius $r_L\ge0.52\,R_{\odot}$ \citep{egg83} for
$M_{\rm NS}\ge1.4\,M_{\odot}$.  This would reduce any ellipsoidal modulation.
If we could double the distance to 1.4~kpc, then the companion
could be an M2 star with $T_{\rm eff}\approx3580$~K and fill its
Roche lobe.  But the difficulty of pulsar wind heating would not be
ameliorated much because $\eta$ would become only slightly smaller,
$\approx0.58$, for the same $T_h$ used above.
However, these conclusions are sensitive to the effective $T_h$ and the
inclination $i$, both of which remain to be measured.

\subsection{Orbital period variations}

\begin{figure}
\begin{center}
\includegraphics[width=0.9\textwidth]{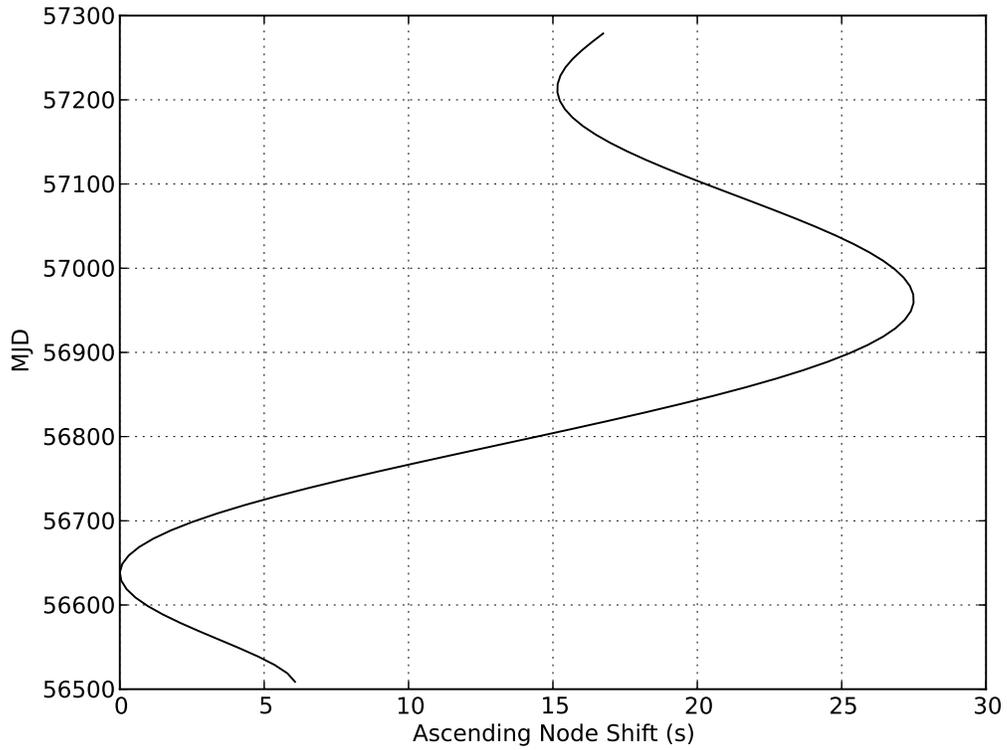}
\caption{The oscillation in the orbital period of PSR~J1048+2339 is shown as a shift in the time of passage through the ascending node vs.\ MJD for the time span of our timing solution in Table~\ref{tab_pulsars}. This is likely caused by changes in the companion's gravitational quadrupole moment.\label{fig_tasc}}
\end{center}
\end{figure}

Figure~\ref{fig_tasc} illustrates the orbital period variations of 
PSR~J1048+2339. Large orbital period variations are known to occur in other 
redback systems such as PSR~J1023+0038 \citep{arc13} and J2339$-$0533
\citep{ple15}.  
Quasi-periodic changes in the gravitational quadrupole
moment of the companion have been proposed to explain observations showing
the orbital period to oscillate \citep{mat83}. 
Several other effects can contribute to orbital period variations: (1)
energy dissipation via gravitational wave emission, (2) mass loss, and
(3) Doppler shifts due to the Shklovskii effect \citep{Shklovskii70}
or Galactic rotation. However, they would result in a monotonic change
in $P_{\rm orb}$ and therefore cannot adequately explain the observed
orbital period variations. 

\cite{Applegate92} relates oscillations in the gravitational quadrupole
moment of a non-degenerate companion star to a cycle of magnetic activity 
similar to the 11-year Solar cycle. The changing magnetic field generates
a variable torque between the convective outer layer and radiative inner
layer of the star. The torque causes angular momentum to be exchanged between
the layers in a quasi-periodic manner, causing changes in the differential
rotation between the layers and therefore in the shape of the star. 
Spin-up of the outer layer corresponds to the star becoming more oblate and to
an increase in the gravitational quadrupole moment. Since the total angular
momentum in the binary system is conserved, this causes a decrease in the
orbital period. \cite{ple15} give an overview of the literature and apply 
the Applegate model to the orbital period variations of PSR~J2339$-$0533. We note that since this model relies on convective and radiative layers being present and stars with mass $\lesssim 0.3$~\msun\ are fully convective, it implies that the companion of PSR~J1048+2339 is more massive than the minimum companion mass $M_{\rm c,min} = 0.3$~\msun and the orbital inclination is $< 90\degrees$.  

Following the same procedure as \cite{ple15}, we obtain a modulation of $P_{\rm orb}$ with a period of $P_{\rm mod} = 1.58$ years and amplitude $A = 1.44 \times 10^{-11}$~s$^{-1}$. Using $M_{\rm c,min} = 0.3$~\msun, and estimated companion radius $R_{\rm c} \approx 0.35$~\rsun~(Section~\ref{sec:companion}), we derive the change in the luminosity of the companion that would account for the energy needed for angular momentum transfer, $\Delta L = 9.44 \times 10^{27}$~erg~s$^{-1}$ (Eqn. 13 from \citealt{ple15}). Based on the estimated companion effective temperature $T_{\rm eff} \approx 3350$~K~(Section~\ref{sec:companion}), its total luminosity is $L_{\rm c} = 5.32 \times 10^{31}$~erg~s$^{-1}$. Since the energy budget for the angular momentum transfer is only $0.02\%$ of the companion's intrinsic luminosity, the companion has sufficient energy output to generate the orbital period fluctuations even if the orbital inclination (assumed to be 90\degrees) is overestimated, or $M_{\rm c}$ or $T_{\rm eff}$ are underestimated. 

\section{Conclusions}

The discovery of the redback pulsar J1048+2339 in a {\em Fermi} unassociated source and the interesting phenomenology in this binary system add a valuable data point for studies of this still not very well understood class of pulsars. Our radio observations show extensive eclipses around superior conjunction as well as sporadic ``mini-eclipses'' at other orbital phases and an occasional long-term absence of pulsed radio emission even near inferior conjunction. This indicates that the ionized gas environment in the system is very dynamic and changes significantly on time scales of hours to days. The radio timing solution that best predicts pulse arrival times includes a fixed proper motion from the APOP catalog \citep{APOP} and five orbital period derivatives. A likely explanation of the rapid orbital period variations is quasi-periodic changes in the gravitational quadrupole moment of the companion due to a magnetic torque mechanism similar to that at play in the 11-year Solar cycle. 

PSR~J1048+2339 has a 20th magnitude optical counterpart detected by SDSS, CRTS, PTF, and Pan-STARRS, whose intensity modulation matches the pulsar orbital period. Its magnitude and colors are consistent with those of an M4 main-sequence star on the ``night'' side. While the optical light curve has the expected minimum at $\phi_{\rm orb} \approx 0.25$, corresponding to inferior conjunction of the companion, it does not exhibit a well-defined maximum at $\phi_{\rm orb} \approx 0.75$. This can be explained by heating not directly from the pulsar but from an asymmetric intra-binary shock. Alternatively, the companion magnetic field may be channeling the pulsar wind to specific regions on the companion's surface that get heated preferentially. 

We did not find an X-ray counterpart in an existing short {\em Swift} observation of the field containing PSR~J1048+2339.
The two-year span of the radio ephemeris does not allow a confident detection of gamma-ray pulsations. Due to the rapid orbital period variations, for the gamma-ray timing analysis we can use only {\em Fermi} data taken within the radio ephemeris time limits. However, the H-test value of a potential gamma-ray pulsation detection improves as the span of the radio timing solution grows, indicating that detection of gamma-ray pulsations is likely as we extend and improve the PSR~J1048+2339 timing solution within the next 1--2 years.

J.S.D. was supported by the Chief of Naval Research and a {\em Fermi} Guest Investigator grant. The Arecibo Observatory is operated by SRI International under a cooperative agreement with the National Science Foundation (AST-1100968), and in alliance with Ana G. M\'endez-Universidad Metropolitana, and the Universities Space Research Association. The Green Bank Telescope is part of the National Radio Astronomy Observatory, a facility of the National Science Foundation operated under cooperative agreement by Associated Universities, Inc. 

The CSS survey is funded by the National Aeronautics and Space Administration under Grant No. NNG05GF22G issued through the Science Mission Directorate Near-Earth Objects Observations Program. The CRTS survey is supported by the U.S. National Science Foundation under grant AST-0909182. The Pan-STARRS Surveys (PS1) have been made possible through contributions of the Institute for Astronomy, the University of Hawaii, the Pan-STARRS Project Office, the Max-Planck Society and its participating institutes, the Max Planck Institute for Astronomy, Heidelberg and the Max Planck Institute for Extraterrestrial Physics, Garching, The Johns Hopkins University, Durham University, the University of Edinburgh, Queen's University Belfast, the Harvard-Smithsonian Center for Astrophysics, the Las Cumbres Observatory Global Telescope Network Incorporated, the National Central University of Taiwan, the Space Telescope Science Institute, the National Aeronautics and Space Administration under Grant No. NNX08AR22G issued through the Planetary Science Division of the NASA Science Mission Directorate, the National Science Foundation under Grant No. AST-1238877, the University of Maryland, and Eotvos Lorand University (ELTE) and the Los Alamos National Laboratory. 

The {\em Fermi}-LAT Collaboration acknowledges generous ongoing support from a number of agencies and institutes that have supported both the development and the operation of the LAT as well as scientific data analysis. These include the National Aeronautics and Space Administration (NASA) and the Department of Energy in the United States, the Commissariat \`{a} l'Energie Atomique and the Centre National de la Recherche Scientifique/Institut National de Physique Nucléaire et de Physique des Particules in France, the Agenzia Spaziale Italiana and the Istituto Nazionale di Fisica Nucleare in Italy, the Ministry of Education, Culture, Sports, Science and Technology (MEXT), High Energy Accelerator Research Organization (KEK) and Japan Aerospace Exploration Agency (JAXA) in Japan, and the K. A. Wallenberg Foundation, the Swedish Research Council and the Swedish National Space Board in Sweden. Additional support for science analysis during the operations phase is gratefully acknowledged from the Istituto Nazionale di Astrofisica in Italy and the Centre National d'\'Etudes Spatiales in France.

\begin{deluxetable}{ll}
\tabletypesize{\small}
\tablecolumns{2}
\tablewidth{0pc}
\tablecaption{Timing solution and derived parameters of PSR~J1048+2339\label{tab_pulsars}}
\tablehead{
\colhead{Parameter} & \colhead{Value\tablenotemark{a}}}
\startdata
\cutinhead{Timing Fit Parameters}
Right ascension, $\alpha$ (J2000) & $10^{\rm h}\,48^{\rm m}\,43^{\rm s}\!.4183(2)$\tablenotemark{b} \\
Declination, $\delta$ (J2000) & $+23\degree\,39\amin\,53\asec\!.411(7)$\tablenotemark{b} \\
Proper motion in $\alpha, \mu_\alpha~{\rm cos}~\delta$ (mas~yr$^{-1}$) & $-18.7$\tablenotemark{c} \\
Proper motion in $\delta, \mu_\delta$ (mas~yr$^{-1}$) & $-9.4$\tablenotemark{c} \\
Spin frequency, $f$ (Hz) & 214.35478538736(7) \\
Spin frequency derivative, $\dot{f}$ (Hz~s$^{-1}$) & $-1.380(8) \times 10^{-15}$ \\
Epoch of timing solution (MJD) & 56897.0 \\
Dispersion measure, DM (pc~cm${}^{-3}$) & 16.6544(1)\tablenotemark{d} \\
Time of passage through ascending node, $T_{\rm asc}$ (MJD) & 56637.598177(1) \\
Projected semi-major axis, $x$ (lt-s)\tablenotemark{e} & 0.836122(3) \\
Eccentricity & 0 \\
Orbital frequency, $f_{\rm orb}$ (Hz) & $ 4.6200376(1) \times 10^{-5}$ \\
1st orbital frequency derivative, $f_{\rm orb}^{(1)}$ (Hz~s$^{-1}$) & $-8.30(3) \times 10^{-18}$ \\
2nd orbital frequency derivative, $f_{\rm orb}^{(2)}$ (Hz~s$^{-2}$) & $-1.6(3) \times 10^{-26}$ \\
3rd orbital frequency derivative, $f_{\rm orb}^{(3)}$ (Hz~s$^{-3}$) & $2.19(6) \times 10^{-31}$ \\
4th orbital frequency derivative, $f_{\rm orb}^{(4)}$ (Hz~s$^{-4}$) & $-2.53(6) \times 10^{-38}$ \\
5th orbital frequency derivative, $f_{\rm orb}^{(5)}$ (Hz~s$^{-5}$) & $1.01(2) \times 10^{-46}$ \\
Planetary Ephemeris & DE421 \\
Time Units & TDB \\
Timing span (MJD) & 56508--57285 \\
Number of observation epochs & 26 \\
Number of points in timing fit & 72 \\
Weighted rms post-fit residual ($\mu$s) & 4.2 \\
[-5pt]
\cutinhead{Derived Parameters}
Galactic longitude, $l$ (deg) & 213.17  \\
Galactic latitude, $b$ (deg) & 62.139  \\
Mass function (\msun) & 0.00100 \\
Surface magnetic field, $B$ (G)\tablenotemark{f,g} &  $3.8 \times 10^{8}$ \\
Spin-down luminosity, $\dot{E}$ (erg~s$^{-1}$)\tablenotemark{g,h} & $1.2 \times 10^{34}$ \\
Characteristic age, $\tau_{c}$ (Gyr)\tablenotemark{g,i} & 2.5 \\
Distance, $D$ (kpc)\tablenotemark{j} & 0.7 \\
\enddata
\tablenotetext{a}{Numbers in parentheses are 1-$\sigma$ uncertainties reported by \textsc{Tempo}.}
\tablenotetext{b}{Correlations between the orbital variability parameterization and the astrometric parameters result in a systematic uncertainty in the timing position of 250 mas (see Section~\ref{sec:energetics}).}
\tablenotetext{c}{The APOP catalog \citep{APOP} value was used as fixed parameter.}
\tablenotetext{d}{Determined from fit to 327 MHz data only.}
\tablenotetext{e}{$x = a_p~\sin\ i/c$, where $a_p$ is the pulsar orbit semi-major axis and $i$ is the inclination angle.}
\tablenotetext{f}{$B = 3.2 \times 10^{19}~(P\dot{P})^{1/2}~{\rm G}$, where $P$ is in seconds.}
\tablenotetext{g}{Not corrected for acceleration effects.}
\tablenotetext{h}{$\dot{E} = 4 \pi^2 I \dot{P} / P^3$ and assuming a moment of inertia $I = 10^{45}$~g~cm$^{2}$.}
\tablenotetext{i}{$\tau_{c} \equiv P / 2\dot{P}$.}
\tablenotetext{j}{Based on DM, sky position, and the NE2001 model of ionized gas in the Galaxy \citep{NE2001}.}
\end{deluxetable}

\end{document}